\documentclass[prl,twocolumn,amsmath,showpacs,floatfix]{revtex4}

\usepackage{graphicx}

\newcommand{\benn}{\begin{displaymath}}
\newcommand{\eenn}{\end{displaymath}}

\renewcommand{\vec}[1]{\mbox{\boldmath $#1$}}

\begin{document}
\input epsf

\title {An intrinsic route to melt fracture in polymer extrusion: \\
a weakly nonlinear subcritical  instability of viscoelastic Poiseuille flow}

\author{Bernard Meulenbroek$^{1}$}\thanks{Present address: CWI,
Postbus 94079, 1090 GB Amsterdam, The Netherlands}
\author{Cornelis Storm$^{1}$}\thanks{Present address: Department of Physics
and Astronomy, University of Pennsylvania, Philadelphia, PA 19104, USA}
\author {Volfango Bertola$^{2}$}
\author{Christian Wagner$^{2}$}
\author{Daniel Bonn$^2$}
\author{Wim van Saarloos$^{1,2}$}

\affiliation{$^{1}$Instituut-Lorentz, Universiteit Leiden, Postbus 9506, 2300 RA
Leiden, The Netherlands\\
$^{2}$Laboratoire de Physique Statistique, Ecole Normale
Sup\'erieure, 24 rue Lhomond, 75231 Paris Cedex 05, France
}

\date{\today}

\begin{abstract}
As is well known, the extrusion rate of polymers from a cylindrical
tube or slit (a ``die'')  is in practice limited by the appearance of
``melt fracture'' instabilities which give rise to unwanted
distortions or even fracture
of the extrudate. We present the   results of a weakly nonlinear
analysis which gives evidence for an intrinsic generic route to melt fracture via   a weakly nonlinear
subcritical instability of viscoelastic Poiseuille flow.
 This instability and the onset of associated melt fracture phenomena appear at a fixed
ratio of the elastic stresses to viscous stresses of the polymer solution.

\end{abstract}


\maketitle

\begin{figure}
\begin{center}
  \includegraphics[width=0.75 \linewidth]{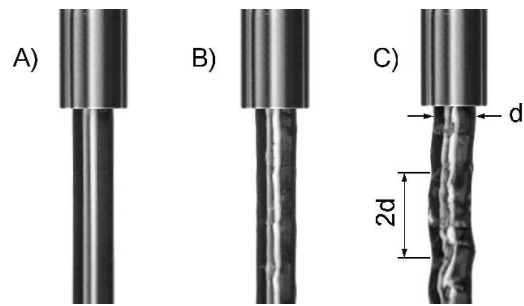}
\end{center}
  \vspace*{-1.5mm}

\caption{Illustration of the surface irregularities that occur in the
extrusion of a polymer fluid from a capillary tube (the wider
structure at the top) in the absence of shark-skin or spurt flow
instabilities. For small speeds,
the surface is smooth {\em (a)}, but beyond some flow speed surface
irregularities develop, {\em (b)} and {\em (c)}.  The wavelength is of
the order of twice the diameter $d$ of the capillary. After \cite{bertola}.}
\label{picture}\label{fig1}
\end{figure}

If a polymeric fiber is produced by extruding it from a so-called die, a
cylindrical tube or planar slit,   the
surface  often exhibits undulations or irregularities at higher flow
rates --- see Fig~\ref{picture}.  For increasing  flow rates  the undulations   become
progressively stronger, so much so that it can eventually  cause
the extrudate  to break --- hence the name {\em melt fracture}. A detailed understanding
 of this phenomenon has
remained elusive for already over 30 years.

Depending on the geometry and type of 
polymer, various types of phenomena seem to occur \cite{denn3,denn4,pahl}. The short
wavelength deformations of the interface often refered to as ``sharkskin'' 
instability appear to originate at the outlet: the extrudate  quasi-periodically
sticks to the outlet, widens, snaps loose and narrows. In the ``spurt-flow'' regime, the extrudate shows
intermittent bands of smooth and irregular surfaces; there is good
evidence that this has to do with a stick-slip phenomenon at the wall
of the die.  In spite of the multitude of possibilities, there is
every reason to believe that when these instabilities are
absent, as in the experiments of Fig.~\ref{picture},  polymer flow
  still exhibits some elusive generic bulk flow instability: 
According to the engineering literature \cite{pahl},   a qualitative  change in the flow behavior  
appears to occur   at a more or less
constant ratio of the normal stress difference 
of the melt over the shear stress for almost any polymer. 
 Isolating this intrinsic mechanism is the main purpose of this paper. 
Our contribution also opens up new avenues for a more general understanding of the rich
variety of complex fluid flow
instabilities, such as 
 polymer turbulence \cite{larson2} or the fact that complex fluids such as granular media 
are prone to instabilities (shear banding) in standard shear configurations.

  The first linear stability analysis of
the flow in the die of a viscoelastic fluid described by the so-called
Oldroyd-B 
constitutive equations \cite{bird}, performed already 
some 25 years ago  \cite{denn1}, showed that the flow was
stable.  Since then it is has been generally accepted that 
 {\em Poiseuille flow of viscoelastic  fluids is linearly stable}. 
The absence 
of a clear {\em linear instability}  explains the focus on other
mechanisms, like those
mentioned above. In this paper we  present the result of an analytical nonlinear amplitude
expansion  which show that
{\em viscoelastic Poiseuille flow exhibits a  weakly nonlinear (or ``subcritical'')
instability} due to normal stress effects; this instability appears to
make  melt fracture phenomena  unavoidable for polymer fluids with 
 normal stress effects, even if care is taken to suppress
instabilities associated with the wall or shape of the die. Recent experiments on polymers whose
rheology  is well captured by the Oldroyd-B  model 
\cite{bertola} confirm our scenario as well as our quantitative predictions.

 Although 
polymer fluids can exhibit all kinds of complicated relaxation
phenomena, the basic feature essentially all polymer fluids (and in
fact most complex fluids) share is the
occurrence of elastic stress effects: when the shear rate is
sufficiently strong that the polymers become stretched by the flow gradients, the
forces along the normals of a little cubical fluid element are
different in different directions, unlike what happens for a Newtonian
fluid where the pressure is isotropic. Microscopically, this effect is
due to the fact that in sufficiently strong shear flows, the polymers get
stretched (like little rubber bands) and acquire a nonisotropic
orientational distribution. Thus they ``pull'' differently on fluid
elements in the direction along the flow and along the shear
gradient. The simplest model to capture this effect (and which hence
has become the working horse of theoretical studies of viscoelastic
polymer flow \cite{bird,larson,shaqfeh2})  is the   so-called
Oldroyd-B model; we study it in the limit of large polymer viscosity,
in which case it is refered to as the Upper Convected Maxwell (UCM) model.
In this regime it is  defined by 
the following constitutive equation for  the shear stress tensor $
\tau$ in terms of the velocity shear 
tensor $\vec{\nabla}\vec{v}$  through   \  
\begin{eqnarray}    
\vec{\vec{\tau }} & + &  \lambda  \left[ {\partial  \vec{\vec{\tau}} }/ {\partial t} +
\vec{v} \cdot \vec{\nabla}     {\vec{\vec{\tau}}}  - (\vec{\nabla } \vec{v})^{\dagger}
\cdot {\vec{\vec{\tau}}} - {\vec{\vec{\tau }}} \cdot ({\vec{\nabla}
}\vec{v})\right]  \nonumber \\
 &  &\hspace*{1cm} =  - \eta   
(\vec{\nabla} \vec{v} + (\vec{\nabla } \vec{v})^{\dagger}),\label{constiteq}
\end{eqnarray}
This constitutive equation  
is characterized by one single relaxation time $\lambda$. The first
two terms between square brackets together constitute the total time
derivative of a fluid element moving with the flow; the other two
nonlinear terms are required by frame-independence, e.g. the fact that
a solid-body rotation of the fluid does not lead to elastic
stress effects. For fluids given by this constitute equation,
the velocity profile in Poiseuille flow is parabolic,
just as in Newtonian fluids.  Moreover, in a cylindrical tube with
radius $R $
and coordinates $(\theta,r,z)$,   the above-mentioned  Weissenberg number
(also called Deborah number)  which is defined as 
\begin{equation}
{\sf Wi} \equiv  \left. \frac{\tau _{rr} - \tau_{zz}}{\tau_{rz}}\right|_{\text{wall}}, \label{weissdef}
\end{equation}
becomes, upon following the usual convention to denote the shear rate
$\partial v^{\text{unp}}_z/ \partial r|_{\text{wall}}$ by $\dot{\gamma}$,   
\begin{equation}
{\sf Wi} =  2 \lambda  \dot{\gamma}
= 4 v_{\text{max}} \lambda/R.  
\end{equation}
Note that the Weissenberg number  defined in (\ref{weissdef}) is
indeed the ratio of the so-called first normal stress difference
$\tau_{rr}-\tau_{zz}$ at the wall
over the  shear stress at the wall. Newtonian fluids are isotropic and
hence the stress difference is zero; for Oldroyd-B model fluids, the
normal stress difference $\tau_{rr}-\tau_{zz} \sim \dot{\gamma}^2$ \cite{bird} while the shear
stress is linear, $\tau_{rz}\sim \dot{\gamma}$;  as a result
${\sf Wi}$ is simply
linear in $\dot{\gamma}$.

\begin{figure}
\begin{center}
 \includegraphics[width=0.6 \linewidth]{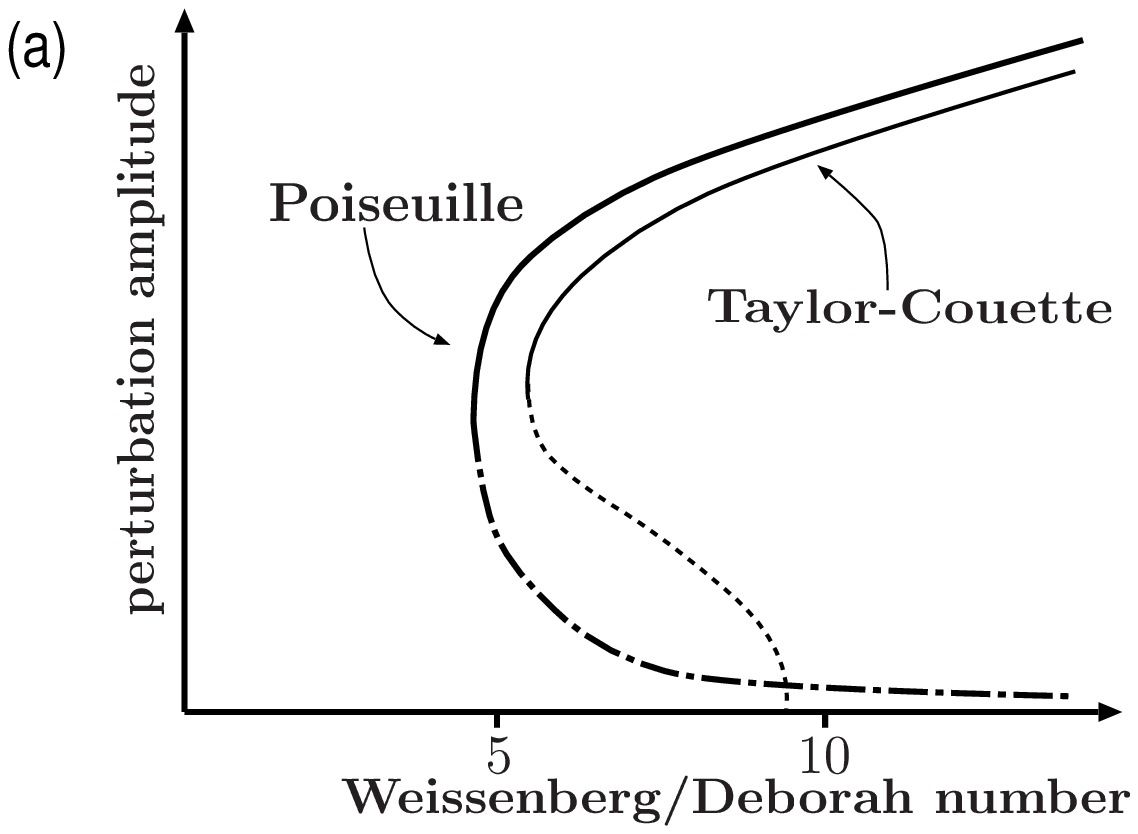}
  \includegraphics[width=0.7\linewidth]{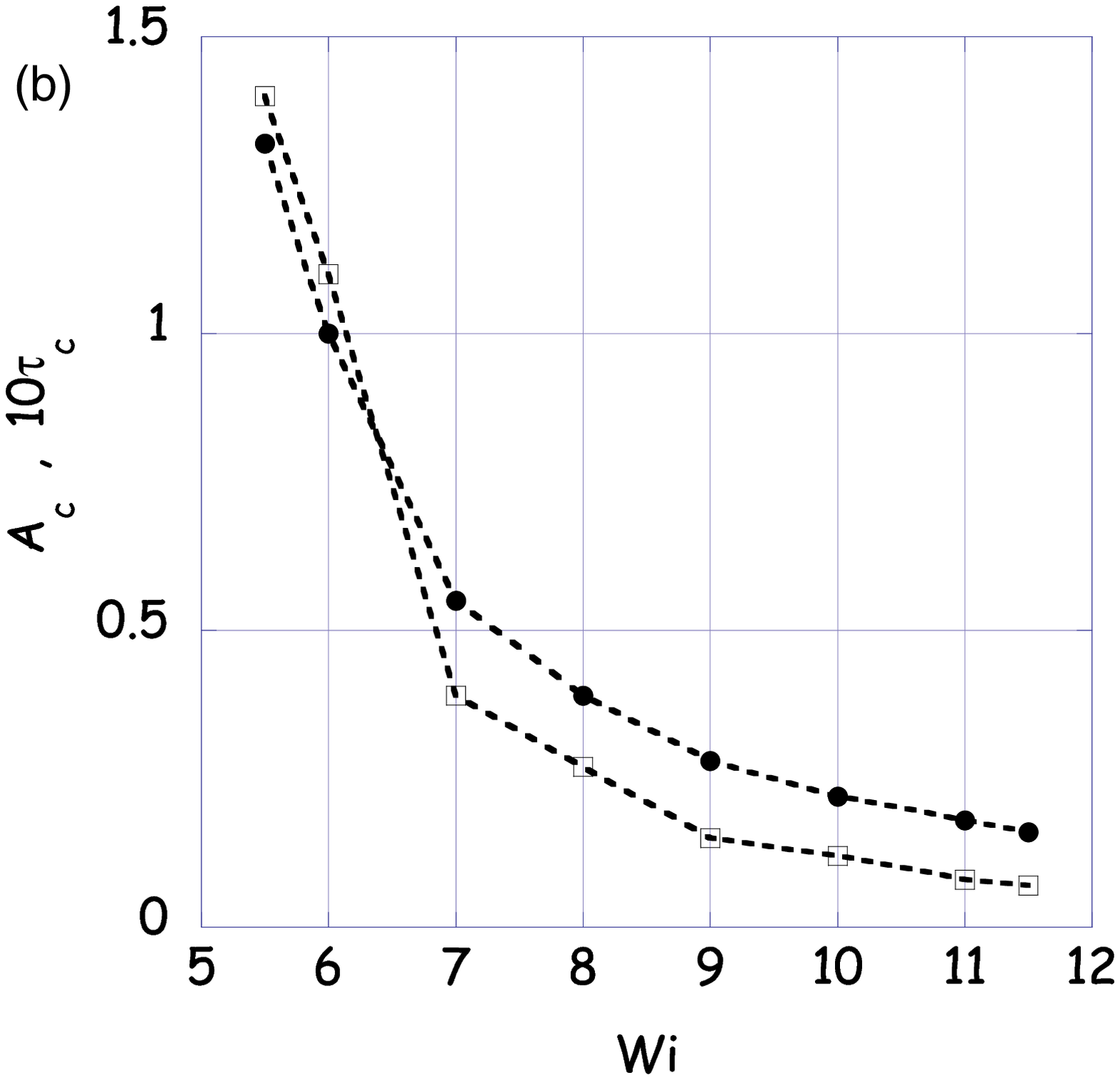}
\end{center}
  \vspace*{-3mm}

\caption{Summary of our main result for the stability of viscoelastic
Poiseuille flow in a cylinder in the zero Reynolds number limit. {\em (a)} 
Semi-quantitative sketch of the bifurcation scenario.  See text. 
{\em (b)} Our results for the threshold values of the amplitude beyond
which the flow is unstable; these curves  correspond to the dashed branch in {\em
(a)}. Dots indicate the critical value of the shear rate at the wall
(normalized to the shear in the unperturbed case), squares the
critical value of the relative shear stress perturbation, multiplied by
10. }
\label{fig2}
\end{figure}

The weakly nonlinear instability scenario that our calculations imply
is illustrated in Fig.~\ref{fig2}a. Along the vertical axis we plot
the relative perturbation of the basic flow profile (e.g. the relative
shear perturbation at the wall, or the relative shear stress
perturbation at the wall) as a function of ${\sf Wi}$. 
Perturbations of the flow whose amplitude is bigger than the value
given by the thick dashed line are unstable; their amplitude grows to
some nonlinear value indicated by the full line. This line denotes
 nontrivial flow behavior with a characteristic
wavelength. Note that the dashed 
line never touches the horizontal axis, in agreement with the fact
that Poiseuille flow is always linearly stable \cite{denn1}, although the threshold
amplitude becomes very small for large ${\sf Wi}$. Furthermore, note that
the full line 
merges with the unstable (dashed) branch (this is called a saddle-node
bifurcation point in technical terms); below the corresponding value
${\sf Wi}_{\text{c}}$, Poiseuille flow is nonlinearly stable to perturbations of any
amplitude. In this paper, we focus on our results for the cylindrical
die, since this is the most relevant case. For slits similar results
are obtained \cite{meulenbroek}.
 
Fig.~\ref{fig2}b shows our analytical  results for the (dashed) branch that marks
the threshold amplitude beyond which the flow us unstable.  As
described in more detail below, our analysis is based on an expansion
to third order in the perturbation amplitude; beyond some critical
value ${\sf Wi}_{\text{c}} \approx 5$, we find that the cubic terms in our expansion
lead to an instability, and from this we are able to calculate the
threshold value. Clearly, our results are fully consistent with the
scenario of Fig.~\ref{fig2}a: the flow is stable to arbitrarily small
(linear) perturbations, but perturbations as small as a few percent in
the shear stress at the wall already render the flow unstable for
values of ${\sf Wi}$ around 10. The critical value of ${\sf  Wi}_{\text{c}}$
we estimate from our
calculations  is also in agreement with the very recent precise experimental
data  on a model UCM fluid \cite{bertola,note1},  and with the range where  
 long  polymers   have been reported to show a   change in the flow behavior \cite{pahl}. 

Actually,   a number of observations  already made us
believe that the scenario summarized in Fig.\ref{fig2} was a viable
one before we embarked on our analysis:
 {\em (i)} For increasing
${\sf Wi}$, the linear stability actually becomes arbitrarily weak (the
damping of the linear modes decreases as $1/{\sf Wi}$). {\em (ii)}
In the zero Reynolds number limit, the flow of an Oldroyd-B fluid in a
Taylor Couette cell (two concentric rotating cylinders) is linearly
unstable above some well-defined value of the Deborah number,
which is analogous to the Weissenberg number \cite{larson,shaqfeh2}. 
This linear instability  is due to the fact that ``hoop stresses''
generally make flow along curved streamlines unstable
\cite{larson,pakdel}.  However, recent experimental investigations
\cite{steinberg2} 
have clearly demonstrated that the instability is ``subcritical'': as
sketched qualitatively in Fig.~\ref{fig2}a, the nonlinear flow branch
corresponding to roll-type patterns in the Taylor-Couette cell has
been shown \cite{steinberg2} to extend down to about half the critical
value where the linear instability occurs (the point where the dotted
line intersects the 
horizontal axis). The subcritical character of the instability has
been argued not to depend on the curvature of the streamlines which
causes the linear instability in the
Taylor-Couette cell. Thus, it is reasonable to assume that both in
viscoelastic Poiseuille flow and in
viscoelastic Taylor-Couette flow there is a subcritical instability,
and that the only essential difference is that in the first case the
dashed branch never intersects the horizontal axis, while in the
second case it  does. {\em (iii)} The transition to turbulence for
Poiseuille flow of Newtonian fluids is also subcritical
\cite{pekeris}. 
\begin{figure}
\begin{center}
  \includegraphics[width=0.7 \linewidth]{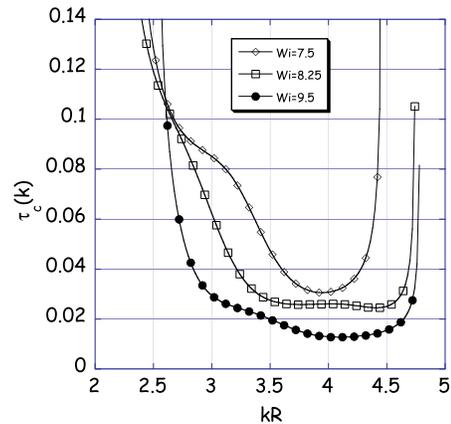}
\end{center}
  \vspace*{-3mm}

\caption{
Values of the  critical  shear stress amplitude $\tau_c(k)$, the ratio
of the perturbation in the shear stress $\tau_{rz}$ (normalized to  
the value $\tau_{rz}^{\text{unp}}$ of the unperturbed state) at the  
wall above which the perturbation is nonlinearly unstable, as a
function of the wavenumber $k$ of the perturbation in the
$z$-direction. }
\label{fig3}
\end{figure}

Returning to our  analysis, it is based on deriving the first
nontrivial nonlinear term in  an
amplitude expansion for  the perturbation to the velocity and shear
stress fields of Poiseuille flow of an UCM fluid. Conceptually,
this method goes back to the classic work \cite{pekeris} in
which the subcritical nature of the transition to turbulence of
ordinary fluids was established; since then, amplitude expansions have
become a standard tool in the field of nonequilibrium pattern
formation  \cite{ch}.  
We write the constitutive equation
(\ref{constiteq}) and the  Navier-Stokes equation (which in the zero
Reynolds number limit is linear) in
cylindrical coordinates and then study the evolution of the amplitude
$A$ of a perturbation $ A (\delta v_{r}, \delta v_{z}, \delta
\tau_{\theta \theta}, \delta \tau_{rr}, \delta \tau_{rz}, \delta
\tau_{zz} ) e^{ikz-i \omega t}$, where the vector is normalized such that the shear
perturbation $\partial  \delta v_z / \partial r$ at the wall equals
$A$ \cite{note2}.  To first nontrivial order
we then obtain an equation of the form
\begin{equation}
d A /dt = - i \omega (k)  A + c_3 |A|^2 A + \cdots \label{cubic}
\end{equation}
To linear order in $A$ this equation simply reproduces the 
term  $\omega(k)$ of the dispersion relation of a single mode
$e^{ikz-i\omega t}$;  this
term is already contained in the old analysis of
\cite{denn1}. In particular, since we know that every mode $k$ is
linearly stable, $\mbox{Im}\, \omega(k) <0$ for all $k$.

The crucial new feature  of our analysis consists of calculating the coefficient
$c_3$ explicitly --- although this is  technically demanding, the
analysis  is standard and conceptually straightforward. It will
therefore be discussed elsewhere \cite{meulenbroek}. 
In particular the real part of $c_3$ is of
importance for determining whether or not the flow is nonlinearly 
unstable: if $\mbox{Re}\, c_3 <0$, then the nonlinear terms increase the 
damping of the amplitude and the unperturbed state is, within this
approximation, also nonlinearly stable. On the other hand, if $\mbox{Re}\, c_3
>0$, then the nonlinear term 
promotes the growth of the amplitude, and in particular amplitudes
\begin{equation}
|A| > A_c  = \sqrt{\frac{  \mbox{Im} \, \omega(k)}{\mbox{Re}\, c_3}} \label{criticalamp}
\end{equation}
grow without bound. Hence, in this approximation $A_c$ 
constitutes the {\em critical amplitude beyond which the flow is
nonlinearly unstable}.  The results presented in Fig.~\ref{fig2}b  for
$A_c$  are obtained directly from our results for the coefficient
$c_3$.  In dimensionless units and with our normalization,  $A_c$
immediately yields the relative shear rate perturbation at the wall,
necessary to make the flow unstable. With a numerical factor that we
obtain  from our analysis, $A_c$ can be converted into a
value for the critical relative shear stress ratio  $\tau_c$, the perturbation
in the shear stress beyond which the flow is unstable.

As we already
discussed in connection with Fig~\ref{fig2}b, for ${\sf Wi} \gtrsim 5$, we
find that there is some range of wavenumbers $k$ for which $\mbox{Re}\, c_3 >0$,
and hence for which the Poiseuille profile is nonlinearly unstable.
In Fig.~\ref{fig3} we show for three values of ${\sf Wi}$ the relative critical shear
stress amplitude $\tau_c(k)$ beyond which the flow is unstable, as a function of
the wavenumber $k$. The further one gets above the value ${\sf Wi}_{\text{c}}$ where
the instability sets in, the wider the band of $k$-values is where we
find an instability; the values plotted in Fig.~\ref{fig2}b for $\tau_c$ correspond
to the minimum values of the curves in Fig.~\ref{fig3}, but as this
figure shows for larger values of ${\sf Wi}$, the bands are very flat so
that the strength of the instability does not appear to depend very
sensitively on $k$ within the unstable band. 

\begin{figure}
\begin{center}
  \includegraphics[width=0.7 \linewidth]{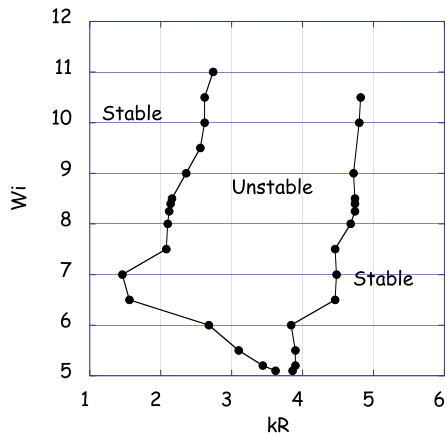}
\end{center}
  \vspace*{-3mm}

\caption{Plot of the width of the $k$-band where the corresponding modes render the 
basic Poiseuille flow profile unstable.}
\label{fig4}
\end{figure}

In Fig.~\ref{fig4} we plot the band in which modes are nonlinearly unstable beyond some
threshold amplitude.  It is
important to realize that $k$ in this figure is the wavenumber of the modulation of
the Poiseuille flow {\em inside} the tube: Since the flow
changes upon exiting the tube, this is not necessarily the same as
the wavenumber of the induced roughness modulation outside the die, as in
Fig.~\ref{fig1}c, but the two can be connected by equating the
{\em frequency} of the modulations measured at a fixed position.  Our numerical results show that the
mode with $kR \approx 3.75$ which according to Fig.~\ref{fig4} emerges at
${\sf Wi} = {\sf Wi}_{\text{c}}\approx 5$ moves with a velocity of about $v
\approx 1.9\, v_{\text{av}}$ where $v_{\text{av}}$ is the average flow
velocity. If the nonlinearities do not change this speed to much, this
suggests a frequency $f$ at onset of about  $3.75\, v/(2\pi R) \approx
1.13\, v_{\text{av}}/R$. Further above threshold our amplitude expansion to cubic order neither
yields
the speed of the finite-amplitude modulated stress pattern nor the
selected wavenumber. However,
it is reasonable to assume that in the saturated nonlinear regime, the
pattern moves with speed $v_{\text{av}}$ in the tube.
From the band of unstable wavenumbers shown in Fig.~\ref{fig4}, we
then conclude that the frequency of the extrudate modulations should
then lie in the range  $0.3 \, v_{\text{av}}/R \lesssim f \lesssim 0.72\,
v_{\text{av}}/R $.

The fact that the critical Weissenberg
number ${\sf Wi}_{\text{c}} \approx 5$ which we find is in  good agreement with
the transition value   noted empirically \cite{pahl}
  is already a strong indication that the
instability which we have identified lies at its origin. Further
independent evidence comes from a series of new experiments which we
have performed on a range of PVA Borax polymer solutions, whose
rheological behavior is well described by the Oldroyd-B/UCM  model (constant
shear viscosity, normal stress proportional to the shear rate squared).
Our scenario implies that the first transition to melt fracture should
be hysteretic, i.e. occur at a higher value of ${\sf Wi}$ upon increasing
the flow rate than upon decreasing the flow rate; this is
 indeed observed in the experiments. Likewise, the
wavelength of the roughness modulations is in good agreement with the
theoretical analysis. We refer to \cite{bertola} for details. 

In conclusion, we shown that normal stress effects drive viscoelastic Poiseuille flow 
unstable. We speculate that  many nontrivial  flow patterns of complex fluids 
that emerge under simple shear
 have a similar origin.




\begin{thebibliography}{99}


\bibitem{denn3} M. M. Denn, {\em Issues in viscoelastic fluid
mechanics}, Annu. Rev. Fluid Mech. {\bf 13}, 13 (1990). 

\bibitem{denn4} M. M. Denn,  Annu. Rev. Fluid Mech. {\bf 33}, 265 (2001).

\bibitem{pahl} M. Pahl, W. Gleissle and H.-M. Laun, {\em Praktische
Rheologie der Kunststoffe und Elastomere} (VDI verlag, 1991).



\bibitem{bird} R. B. Bird, R. C. Armstrong and O. Hassager {\em
    Dynamics of Polymeric Liquids} (Wiley, New York, 1987).


\bibitem{denn1}
T. C. Ho and M. M. Denn, J. Non-Newtonian Fluid Mechanics {\bf 3},
179 (1978).



\bibitem{larson2} G. V. Vinogradov and V. N. Manin, Kolloid Z. {\bf
201}, 93 (1965); A. Groisman and V. Steinberg, Nature {\bf 405}, 53 (2000); R. G. Larson,
Nature {\bf 405}, 27 (2000).

\bibitem{bertola} V. Bertola,   B. Meulenbroek, C. Wagner, C. Storm,
W. van Saarloos, and D. Bonn (unpublished).


\bibitem{larson} R. G. Larson, E. S. G. Shaqfeh and S. J. Muller, J. Fluid
Mech. {\bf 218}, 573  (1990).

\bibitem{shaqfeh2} E. S. G. Shaqfeh, Annu. Rev. Fluid Mech. {\bf 28}, 129
(1996).




\bibitem{meulenbroek} B. Meulenbroek, C. Storm and W. van Saarloos,  submitted to Phys. Fluids.

\bibitem{note1} We stress that our amplitude expansion to cubic order  is selfconsistent  
for the determining the nonlinear instability threshold for large
${\sf Wi}$,
where  the threshold values are small. Near the
saddle-node at ${\sf Wi}_{\text{c}}$, higher order terms are expected to be important
quantitatively.  

\bibitem{pakdel} P. Pakdel and G. H. MacKinley, Phys. Rev. Lett. {\bf 77}, 2459
(1996).

\bibitem{steinberg2} A. Groisman and V. Steinberg, Phys. Fluids {\bf 10}, 2451 (1998).

\bibitem{pekeris} 
P. Huerre and M. Rossi, in: {\em Hydrodynamics and
  Nonlinear Instabilities}, C. Godr\`eche and P. Manneville,   eds. (Cambridge UP, Cambridge, 1998).


\bibitem{ch} M. C. Cross and P. C. Hohenberg, Rev. Mod. Phys. {\bf
65}, 851 (1993).


\bibitem{note2} We assume no $\theta$-dependence of the modes;  in
spite of this, $\tau_{\theta \theta} \neq 0$ 
in cylindrical coordinates.


\end{thebibliography}
\end{document}